\documentclass[12pt]{article}
\usepackage{epsfig}
\usepackage{amsfonts}

\def\laq{~\raise 0.4ex\hbox{$<$}\kern -0.8em\lower 0.62
ex\hbox{$\sim$}~}
\def\gaq{~\raise 0.4ex\hbox{$>$}\kern -0.7em\lower 0.62
ex\hbox{$\sim$}~}

\def\beq{\begin{equation}}
\def\eeq{\end{equation}}
\def\bea{\begin{eqnarray}}
\def\eea{\end{eqnarray}}

\def \Da {\Delta}

\def \noi {\noindent}

\begin{document}
\begin{titlepage}

\begin{flushright}
BA-TH/810-25\\
%arXiv:08mm.nnnn
\end{flushright}

\vspace{1 cm}

\begin{center}

\LARGE{On some properties of that strange component of Nature called ``time"}

\vspace{1cm}

%\large
\normalsize
{M. Gasperini}

\bigskip
\normalsize

{\sl Dipartimento di Fisica,
Universit\`a di Bari, \\
Via G. Amendola 173, 70126 Bari, Italy\\
and\\
Istituto Nazionale di Fisica Nucleare, Sezione di Bari, Bari, Italy \\
\vspace{0.3cm}
E-mail: {\tt gasperini@ba.infn.it}

%\vspace{0.3cm}
%Submission date: {\rm 10 February 2025}
}

\vspace{1cm}

\begin{abstract}
\noi
Are time-travels possible? is the past still existing? and is the future already existing? We try to give an answer to these an other questions concerning the properties of time and the close connection (but deep physical difference) between  time and space. 
\end{abstract}
\end{center}

\bigskip
\begin{center}
---------------------------------------------\\
\vspace {5 mm}
{\em Essay written for the Gravity Research Foundation}\\
{\em 2025 
Awards for Essays on Gravitation}

\vspace {5 mm}
To appear in: {\bf The European Physical Journal Plus (2025)}

{EPJP-D-25-02914}

\end{center}

\end{titlepage}

\newpage
\parskip 0.2cm

Space and time: after the wonderful of Einstein, and his relativistic description of the world, we are all tempted to put space and time on the same level, implicitly assuming a close physical and conceptual equivalence between these two basic ingredients of Nature. The aim of this short Essay is to try to convince the reader that the notions of space and time describe instead completely different aspects of the physical reality, and that taking into account such a difference is probably unavoidable for future advances in physics.

Let us start with a trivial remark. We can move in space forward and backward, left and right. And in time? The answer can be synthetically summarized as follows.

Time travels? Possible. Traveling back to our own past? Impossible. Put simply, we can leave home, take a trip, go back home and find that the home clock is forward, or back, with respect to our own watch. However, the time shown by the home clock is always {\em later} than the time shown at our departure. 

When we move in space, in fact, we can experience three possible different  situations.
\begin{itemize} 
\item[\textbf{1 -}]  Suppose we leave from a spatial position $x_0$ at a time $t_0$, with our watch exactly synchronized with  the local clocks, and that we come back to the same point $x_0$ after a time interval $\Da t$, according to our watch. We then find that for the local clocks at rest in $x_0$ a longer time interval has passed, $\Da t' > \Da t$. This is the well known ``twin paradox" (see e.g. \cite{1,2}), due to the slowing down of the time flow for moving bodies with respect to those at rest. Compared with our own proper time we are then ``in the future" of the spatial point $x_0$: even if we stop at that position we cannot have any direct experience of an epoch of duration $\Da t' - \Da t$ which for that point has already passed for ever. 

\item[\textbf{2 -}]  Suppose now that we leave from the point $x_0$ at the time $t_0$, synchronized as before with the local clocks, that we come back to $x_0$ after a proper time interval $\Da t>0$, and we find that for the local clocks a shorter time interval has passed, $\Da t' < \Da t$. This is also a physically possible result if our trip is made, at least partially, along a ``closed timelike trajectory" like those possibly produced by strong enough gravitational fields and appropriate topological configurations of space, like black holes and wormholes (see e.g. \cite{3,4}). In that case, and with respect to our own proper time, we landed ``in the past" of the point $x_0$: we are back by a time interval $\Da t- \Da t'$ which in that point has yet to pass, and which we can directly experience if we keep fixed at $x_0$. 

\end{itemize}
However, it is important to stress that the mentioned mechanisms of gravitational and topological type, able to ``accelerate" or to ``decelerate" the proper-time flow of a traveller with respect to that of the clocks at rest, cannot in any case allow the traveller to come back to $x_0$ after a local time interval $\Da t' = t'-t_0 <0$: namely, they do not allow the return at a local epoch $t'$ {\em earlier} than the local departure time $t_0$. 

But the scenario may be different if we consider a travel between two {\em different} points of space. In such a case we are led to a third possible situation. 

\begin{itemize}
\item[\textbf{3 -}] Suppose we start from a spatial position $x_1$ at a time $t_1$ according to our own watch, and that we arrive at a proper time $t_2$ to a different spatial position $x_2$, where the time marked by the local clocks is $t'_2$. It is possible, in principle, that $t'_2 < t_1$, namely that we ended up in a place with a local time {\em prior} to that of our departure: in such a case we may live again epochs that we have already experienced before leaving, but we are experiencing them {\em in a different spatial position}. 

Hence, even if we stop at the point $x_2$, and wait until the local time marks our departure time, i.e. $t'=t_1$, we cannot in any case locally meet and/or directly interact with ourselves (so as, for instance, to prevent us from leaving), because at those times we were in a different spatial position. If on the contrary we keep traveling, leaving from $x_2$ to come back to our initial position $x_1$ (hoping to arrive before our departure), we would be again in one of the two previous cases, and we would arrive to $x_1$ at a local time $t>t_1$, i.e. always later than the time of our initial departure. The same would happen if, instead of moving ourselves from $x_2$, we would try to send signals from the point $x_2$ to the point $x_1$.
\end{itemize}
The above situations, and the above results, are useful and instructive to try to answer some questions about time that are often posed at a philosophical level, but which also corresponds to valid scientific issues. Typical questions like: does the future already exist? Does the past still exists? They are very deep questions in general, but the answer of physics, according to our present knowledge, is rather simple. 

Taking also into account the previous examples we can say that the answer is NO  to both questions, if the different instants of time which we select to define the past and the future are all referred to our own flow of time, marked by our own watch. Or, to put it in a slightly more technical language, if we refer to instants of time all located on the same world-line of a given (unique) observer. 

The answer may be YES to both questions, instead, if the different instants of time to which we refer are located on different world-lines of different observers, independently of the spatial position. For a simple, but concrete example of this effect we can take for instance the previously mentioned twin paradox. For the travelling twin, the twin at rest is older, hence in his future; for the twin at rest, on the contrary, the travelling twin is younger, hence in his past. Nevertheless they can meet together at the same spatial position, where they can hug each other and stay in direct physical contact.

We are thus led to consider a scenario where, at any given point of space, the {\em local} proper time incessantly flows towards the future, producing continuous and unstoppable local changes. We can pass two (or more) times through the same point of space, but the second time we pass we always find that the {\em local} clocks mark a time {\em later} than the local time of our previous passage. 

Such a scenario, which might appear rather simple and natural, actually becomes highly non-trivial in view of the fact that the local flow of time has ``non-universal" properties: time may flow at different speeds in different points of space, and in different physical conditions. As already seen with the previous examples this gives us the possibility, moving in space, to make also a sort of ``time travels" by which  we end up in places and physical configurations corresponding to the past or to the future with respect to our own watch and calendar. 

The different flowing speed of local time is well understood at a classical and macroscopical level, where it corresponds to different physical situations like the relative motion at high velocity (see the previous example {\bf 1}), or the presence of strong gravity fields and suitable spatial topologies (example {\bf 2}). But a locally varying speed of the flow of proper time must (unavoidably) exist also at a microscopic level, as a consequence of the fact that the flowing speed of time, like all other microscopic observable quantities, is continuously ``fluctuating", i.e. may vary from point to point in a random and unpredictable way because of intrinsic quantum effects. 

Such a time fluctuation at the local microscopic level can provide a simple explanation of the so-called ``mean lifetime", the well known property of decaying particles (and, more generally, of unstable quantum states).

Let us recall indeed that elementary particles of the same type (like for instance, electrons, muons, etc) have all exactly the same mass (as experimentally checked with always increasing precision), exactly the same charge (modulo a possible change of sign for the corresponding antiparticle), the same spin, and also the same {\em ``mean lifetime"}. Why not, instead,    the {\em same lifetime}?

The muons, for instance, have a mean lifetime $\tau_\mu \simeq 2.2 \times 10^{-6}$ sec. Let us suppose that we have produced at the initial time $t_0$ a number $N_0$ of identical muons, all with the same mass and charge, which are positioned in our laboratory and suitably isolated from any type of external interaction. We will observe that all muons {\em do not} simultaneously decay after a time interval $\Da t = \tau_\mu$, namely we will not observe that their number decreases in time, for $t>t_0$, like $N(t)= N_0 \,\theta(\tau_\mu-t+t_0)$, where $\theta$ is the Heaviside step function. Actually, we will see that their number decreases in time according to the well known exponential law $N(t)= N_0 e^{-(t-t_0)/\tau_\mu}$. This necessarily implies that there are muons that decay ``earlier" (after a time $\Da t < \tau_\mu$) with respect to an external observer, while other muons decay ``later" (after a time $\Da t > \tau_\mu$).

This might be explained, classically, if some of these muons would move at very high velocity and/or would be immersed in strong gravitational fields and/or would interact with non-trivial spatial topologies. We have seen indeed that such ingredients may crucially affect the  flow of proper time. However, none of such ingredients affects our muons: gravity in our laboratory is weak, and the muons are produced at very low energy and very small velocities.

But if the velocity of each muon is given with a known experimental precision then, according to the quantum uncertainty principle, we cannot exactly known the corresponding position. This implies that the single particles of our group of muons are localized, in general, {\em in different points of space}, and suggests that such a different microscopic localization might be the reason of the difference in the various effective times of decay. 

Indeed,  let us come back to the rather natural assumption that identical particles have not only the same mass, charge, spin, etc, but also have (in proper time) the same  {\em identical lifetime} $\tau$. Let us also assume, however, that at the microscopic level the local flow of proper time fluctuates as a quantum variable, and thus varies from point to point\footnote{The local quantum oscillations of the proper time of a stationary particle have been considered (with different motivations) also in \cite{5}.}. Particles placed at different spatial positions, even if at rest, are thus affected by the different local behavior of  proper time, corresponding to either an accelerated or slowed down local flow and producing, for the external observer, a different effective decaying time for different particles. Note that the explicit fluctuation law of local time is not fully arbitrary, being determined by a statistical condition: its spatial average must reproduce the exponential decreasing of the number of ``still alive" particles, namely it must satisfy the average constraint $\langle dN(t)/dt \rangle=- \langle N(t)/\tau_\mu\rangle$. 

To conclude this short Essay let me recall that, in the context of modern theoretical physics, we often find the statement ``time doesn't exist" (see e.g. \cite{6}). I would say instead that time, interpreted as a peculiar aspect of the physical reality, does exist. Nature shows us that time is an intrinsic property of all points of space. Rather, I would say that what does not exist is the ``space-time"\footnote{I agree that this may be somewhat difficult to be accepted by people who (like me) have been working for many years, and are still working, on general relativity and gravitational theories. But it would be wrong to fully identify the physical reality with the formal models we use for its mathematical description. The spacetime variety is only a mathematical tool useful to describe the physical reality, just like (for instance) the Hilbert space of quantum mechanics.}. Namely, it does not exist a true time coordinate fully equivalent to the coordinates we use for the spatial dimensions, and a four-dimensional geometric variety that we (or, better, Einstein and Minkowski) have invented to provide a simplified formal description of the physical effects of time.  

It is perhaps instructive, at this point, to make an explicit analogy with the case of the electric charge and electromagnetic field. The electric charge undoubtedly exists in Nature, and is a property that the physical systems may have, and by which can interact among themselves. The electromagnetic field, instead, is an our own mathematical construction, useful for a phenomenological description of the forces  between the charged systems. 

In the same way, time exists as an intrinsic property of all physical systems and all points in space, while the time coordinate, as well as the associated space-time variety, is a simple mathematical tool, quite efficient for a quantitative description of all dynamical processes.

Just like the electric charge may be positive, negative or null, also the time has a ``sign": it can flow in one direction (for matter), in the opposite direction (for antimatter), or not flow at all (for light rays and massless particles).

Finally, just like the electromagnetic field (i.e., our formal construction representing the physical effects of charge) is subject to microscopic quantum fluctuations, the time coordinate (introduced to describe the physical effects of time) also must be locally subject to quantum microscopic fluctuations, as confirmed for instance by the statistical decay law of unstable particles.

%%%%%%%%%%%%%%%%%%%%%
%%%%%%%%%%%%%%%%%%%%%%

\section*{Acknowledgements}

Dedicated to the memory of my wife, Patrizia Bolognesi, for her invaluable support and love throughout the time we spent together.

\section*{Data Availability Statement}

No Data associated in the manuscript.

%\vspace{1 cm}

%\newpage


\begin{thebibliography}{99}
\newcommand{\bb}{\bibitem}



\bb{1}R. Resnick, {\em ``Introduction to special relativity"}, John Wiley \& Sons (London, 1968). 

\bb{2}W. Rindler, {\em ``Introduction to special relativity"}, Clarendon Press (Oxford, 1982). 

\bb{3}S. Krasnikov, {\em ``Back-in-time and faster-than-light travel in general relativity"}, Springer Int. Pub. (2018).

\bb{4}F. S. N. Lobo (Editor), {\em ``Wormholes, warp drives and energy conditions"}, Springer Int. Pub. (2018).

\bb{5}Hou. Y. Yau, {\em ``Quantum properties and gravitational field of a proper time oscillator"}, arXiv:0706.0190 [physics.gen-ph].

\bb{6}C. Rovelli, {\em ``And if time doesn't exist?"}, Editions Dunod (2012). 

\end{thebibliography}
\end{document}